\documentclass[twocolumn,tighten]{aastex62}
\usepackage{graphicx}	
\usepackage{amsmath}

\newcommand{\etal}{et al.}

\newcommand\xte{{\it RXTE\/}}

\newcommand\chandra{{\it Chandra}}
\newcommand\xmm{{\it XMM-Newton\/}}

\newcommand\nustar{{\it NuSTAR\/}}

\newcommand\snr{Kes~75}

\newcommand\psr{PSR~J1846$-$0258}

\newcommand\RPP{rotation-powered pulsars}

\def\simlt{\mathrel{\hbox{\rlap{\hbox{\lower4pt\hbox{$\sim$}}}\hbox{$<$}}}}
\def\simgt{\mathrel{\hbox{\rlap{\hbox{\lower4pt\hbox{$\sim$}}}\hbox{$>$}}}}

\submitjournal{ApJ}

\shorttitle{Broadband X-ray Study of Kes~75}
\shortauthors{Gotthelf, Safi-Harb, Straal \& Gelfand}

\begin{document}

\title{X-RAY SPECTROSCOPY OF THE HIGHLY MAGNETIZED PULSAR PSR~J1846--0258, \\ ITS WIND NEBULA AND HOSTING SUPERNOVA REMNANT KES~75} 
\author{E.~V. Gotthelf}
\affil{Columbia Astrophysics Laboratory, Columbia University, 550 West 120th Street, New York, NY 10027-6601, USA}
\author{S. Safi-Harb}
\affil{Department of Physics and Astronomy, University of Manitoba, Winnipeg, MB R3T 2N2, Canada}
\author{S.~M. Straal}
\affil{NYU Abu Dhabi, PO Box 129188, Abu Dhabi, United Arab Emirates}
\affiliation{Center for Astro, Particle, and Planetary Physics (CAP$^3$), NYU Abu Dhabi, PO Box 129188, Abu Dhabi, United Arab Emirates}
\author{J.~D. Gelfand}
\affil{NYU Abu Dhabi, PO Box 129188, Abu Dhabi, United Arab Emirates}
\affiliation{Center for Astro, Particle, and Planetary Physics (CAP$^3$), NYU Abu Dhabi, PO Box 129188, Abu Dhabi, United Arab Emirates}
\affiliation{Center for Cosmology and Particle Physics (CCPP, Affiliate), New York University, 726 Broadway, Room 958, New York, NY 10003}

\correspondingauthor{E. V. Gotthelf; eric@astro.columbia.edu}

\begin{abstract} 
  We present broad-band X-ray spectroscopy of the energetic components
  that make up the supernova remnant (SNR) Kesteven 75 using
  concurrent 2017 Aug 17-20 \xmm\ and \nustar\ observations, during
  which the pulsar \psr\ is found to be in the quiescent state. The
  young remnant hosts a bright pulsar wind nebula powered by the
  highly-energetic ($\dot E = 8.1 \times10^{36}$~erg~s$^{-1}$)
  isolated, rotation-powered pulsar, with a spin-down age of only
  $P/2\dot P \sim 728$~yr.  Its inferred magnetic field ($B_s =
  4.9\times 10^{13}$~G) is the largest known for these objects, and is
  likely responsible for intervals of flare and burst activity,
  suggesting a transition between/to a magnetar state.  The pulsed
  emission from \psr\ is well-characterized in the 2$-$50~keV range by
  a power-law model with photon index $\Gamma_{\rm PSR} = 1.24\pm0.09$
  and a 2$-$10~keV unabsorbed flux of $(2.3\pm0.4)\times
  10^{-12}$~erg~s$^{-1}$~cm$^{-2}$. We find no evidence for an
  additional non-thermal component above $10$~keV in the current
  state, as would be typical for a magnetar.  Compared to the
  \chandra\ pulsar spectrum, the intrinsic pulsed fraction is
  $71\pm16$\% in 2$-$10~keV band.  A power-law spectrum for the PWN
  yields $\Gamma_{\rm PWN} = 2.03\pm0.02$ in the 1$-$55 keV band, with
  no evidence of curvature in this range, and a 2$-$10~keV unabsorbed
  flux $(2.13\pm0.02)\times 10^{-11}$~erg~s$^{-1}$~cm$^{-2}$.  The
  \nustar\ data reveal evidence for a hard X-ray component dominating
  the SNR spectrum above 10 keV  which we attribute to a
  dust-scattered PWN component. We model the dynamical and radiative
  evolution of the Kes~75 system to estimate the
  birth properties of the neutron star, the energetics of its
  progenitor, and properties of the PWN. This suggests that the
  progenitor of Kes~75 was originally in a binary system which
  transferred most its mass to a companion before exploding.
\end{abstract}

\keywords{ISM: individual (Kes~75) --- pulsars: individual (\psr) --- stars: neutron}

\section{Introduction} \label{sec:intro}

To date over 2000 `ordinary' rotation-powered pulsars (RPPs) have been
discovered, nearly all as radio pulsars. Their beamed emission is
powered by the rotational energy loss from a radiating magnetic dipole
of the neutron star (NS) as it gradually slows down \citep{sha83}.
Young, energetic isolated pulsars often display radio and/or X-ray
pulsar wind nebulae (PWN). It is the conversion of rotational energy
into electromagnetic radiation and a particle wind that is thought to
energize these synchrotron nebulae \citep{gaensler06}.

In contrast, magnetars \citep[e.g.,][]{tur15} are also young, isolated
NSs but they typically lack the persistent radio emission and the
synchrotron nebulae of the \RPP. In their quiescent state, these
slowly rotating pulsars ($2-12$~s) emit uniquely in the X-ray band.
Most notably, their thermal X-ray emission far exceeds their
rotational kinetic energy loss rate ($\dot{E}$) and are thought to be
powered, instead, by the decay of their enormous magnetic field, above
the quantum critical value of $B_{\rm QED}\equiv m_e^2c^3/e\hbar = 4.4
\times 10^{13}$~G \citep{dun92}.

Recent exceptions to the defining properties of both the \RPP\ and the
magnetars are shedding new light on their evolution and emission
mechanisms: 1) over the last few years, a growing number of ``low
B-field'' magnetars have been detected, whose magnetic fields border
those of the \RPP\ \citep[e.g.,][]{rea14} 2) the so-called transient
magnetars which are seen to display intermittent radio emission during
their outbursts \citep[e.g.,][]{cam08}, 3) recent evidence for the
first detection of a wind nebula around a magnetar \citep[{\it Swift}
J1834.9$-$0846;][]{you16} and 4) the discovery of
variability in a PWN surrounding the rotation-powered, high-B radio
pulsar J1119--6127, that has shown a magnetar-like behavior
\cite{blu17}.

A remarkable new result is the discovery of magnetar-like millisec
bursts from \psr\ in \snr\ \citep{gav08,kum08, ng08}. This highly
energetic ($\dot E = 8.3 \times 10^{36}$~erg~s$^{-1}$), 325~ms X-ray
rotation-powered pulsar \citep{got00}, with its bright wind nebula, is
located within the core of the young SNR \snr\
\citep{hel03,su09,tem12}. The pulsar was caught serendipitously in
2006 in a flaring state, with a notably softening of its spectrum and
some PWN morphology change. And most recently, 14 years later, renewed
activity is reported from the pulsar \citep{kim20}.  However, the
energetics and spectral properties of this pulsar otherwise strongly
distinguish it from a magnetar. We might be catching a unique and rare
evolutionary state, possibly observing breakout from a buried
magnetar-strength magnetic field (see \citealt{hal10}).

\begin{figure*}[t]
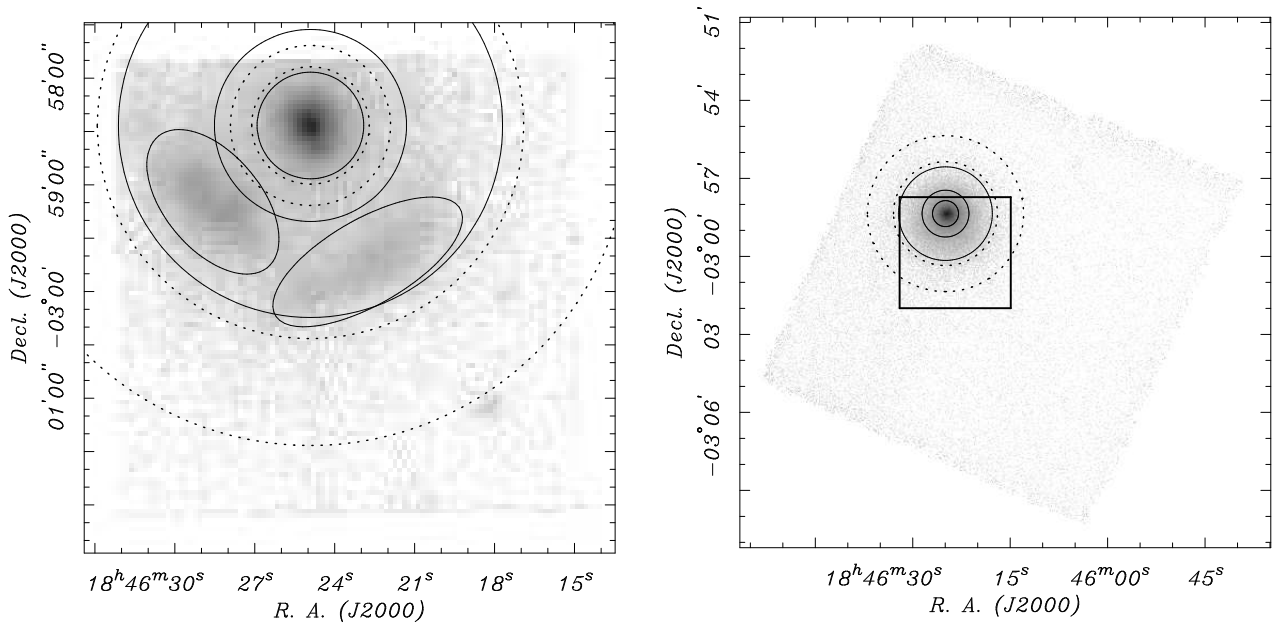

\vspace{0.1in}
\centerline{
\hfill
\includegraphics[angle=270,width=0.45\linewidth]{./kes75_snr_xmm_epn_1000_10000_img_corr_fixed.eps}
\hfill
\includegraphics[angle=270,width=0.45\linewidth]{./nu40301004002_35_460_corr.eps}
\hfill
}
\caption{Low resolution X-ray images of \snr\ scaled logarithmically
  to highlight faint emission. Shown are the spectral extraction regions for the source ({\it solid circles}) and background ({\it   dotted circles}) as described in the text.  Left --- The \xmm\ EPIC pn small window mode image in the 1$-$10~keV band. The SNR emission is highly cut off below 1~keV. Right --- The \nustar\ 3$-$79~keV image.  The pulsar and PWN are unresolved and overlap the much fainter SNR emission in this band. The EPIC pn field of view is overlaid on the \nustar\ image (solid box).}
\label{fig:images}
\end{figure*}

We have obtained new, concurrent \xmm\ and \nustar\ observations of
the Kes~75 system in order to better study the spectral properties of
its components, to infer their interactions and evolution.  In \S2, we
describe these data sets, analyzed in conjunction with a recent
archival \chandra\ data.  We find that a simple power-law model is
sufficient to characterize the spectrum of the pulsar, with no
evidence for a flatter spectral component above $\sim10$~keV, a
characteristic of the magnetars. The spectrum of the PWN, also
well-modeled by a single power-law, shows no evidence of curvature. We
measure a dust scattered halo that manifest as a hard $>10$~keV
spectral component for the SNR spectrum.

Based on these spectral results, in \S3 we apply a dynamical and
radiative evolution model to spectral energy distribution (SED) of the
Kes~75 system. Finally, in \S4, we discuss the implications of our
broad-band spectral and evolution modeling. The discovery of
magnetar-like activity from the young rotation-powered pulsar offers a
unique opportunity to study the connection between distinct nature of
\RPP\ and magnetars, their birth and early evolution.

\section {X-ray Observations and Analysis}

\subsection {\nustar}

We observed \snr\ with \nustar\ 2017 Aug~17 as part of the AO3 Guest
Observer programs.  The nominal 100~ks exposure started at 18:56:09~UT
and lasted for 208~ks, with Earth block accounting for periodic
orbital time gaps.  \nustar\ consists of two co-aligned X-ray
telescopes, with corresponding focal plane detector modules FPMA and
FPMB, each of which is composed of a $2\times2$-node CdZnTe sensor
array \citep{Harrison2013}.  These are sensitive to X-rays in the
3$-$79~keV band, with a characteristic spectral resolution of 400~eV
FWHM at 10~keV. The multi-nested foil mirrors provide
$18^{\prime\prime}$ FWHM ($58^{\prime\prime}$ HPD) imaging resolution
over a $12\farcm2\times 12\farcm2$ field-of-view (FoV)
\citep{Harrison2013}.  The nominal timing accuracy of \nustar\ is
$\sim$2~ms rms, after correcting for drift of the on-board clock, with
the absolute timescale shown to be better than $<$$3$~ms
\citep{Mori14, Madsen15}.  This is more than sufficient to resolve the
signal from \psr.

\nustar\ data were processed and analyzed using {\tt FTOOLS}
09May2016\_V6.19 ({\tt NUSTARDAS} 14Apr16\_V1.6.0) with \nustar\
Calibration Database (CALDB) files of 2016 July~6.  The resulting
dataset provides a total of 95~ks of net good exposure time. For all
subsequent analysis we merged data from the two FPM detectors.  As
shown in Figure~\ref{fig:images}, the small diameter SNR \snr\ was
fully imaged on the FPM detectors, centered on node-0. The pulsar is
unresolved from the PWN emission, and its contribution is taking into
account in the spectral analysis, using \chandra\ data.

\subsection {\xmm}

We also obtained a shorter, but uninterrupted 51.4 ks \xmm\
observation of \snr\ on 2017 Aug~19 starting at 14:28:17~UT,
overlapping the end of the \nustar\ observation, $1.81$~days from its
start. The European Photon Imaging Camera (EPIC) on-board \xmm\
consists of three detectors, the EPIC pn detector \citep{Struder01}
and EPIC MOS1 and MOS2 \citep{Turner01}.  These sit at the focal plane
of co-aligned multi-nested foil mirrors with an on-axis point spread
function with FWHM of $\sim$$12\farcs5$ and $\sim$$4\farcs3$ at
1.5~keV, for the pn and MOS, respectively. The EPIC detectors are
sensitive to X-rays in the 0.15$-$12~keV range with moderate energy
resolution of $E/\Delta E({\rm pn}) \sim $20$-$50.

The EPIC pn data were obtained in {\tt PrimeSmallWindow} mode
($4\farcm3\times4\farcm4$), with an increased time resolution of 6~ms,
sufficient to phase-resolve \psr, at the expense of a large 29\%
deadtime. The pulsar, PWN, and the main clumps of the SNR are imaged
in the FoV (see Fig~\ref{fig:images}).  To resolve possible rapid
bursts from the pulsar we operated EPIC MOS2 in {\tt FastUncompressed}
mode, which offers 1.5~ms time resolution with 1D-imaging on the
central EPIC CCD, at the expense of an increase in the background
component. The MOS1 data were acquired in $1\farcm8\times1\farcm8$
{\tt PrimePartialW2} small-window mode on the central CCD with the PWN
just filling the reduced FoV. The time resolution in this mode is
300~ms, useful for searching for slower flares but insufficient to
resolved \psr.

Data were reduced and analyzed using the Standard Analysis Software
(SAS) v.15 with the most up-to-date calibration files. After filtering
out background flares we obtained usable live time of 50.1/36.1~ks,
for the MOS/pn data.

\subsection{Archival \chandra\ data}

To correct and verify our \xmm\ analysis of \snr, we used archival
\chandra\ data to spatially resolve in the same band-pass the pulsar,
PWN, and SNR components and obtain the most accurate fluxes. Of the
many \chandra\ observations of \snr\ obtained over the years, we
select for this work ObsID \#18030, a deep observation acquired
closest in time to our concurrent \xmm\ and \nustar\ data set. This
observation was carried out with the Advanced CCD Imaging Spectrometer
(ACIS) on 2016 June 8 with an exposure time of 86~ks. The data were
reprocessed and cleaned using standard CIAO v4.10 routines, resulting
in an effective exposure time of 84.9~ks. The \chandra\ count rate
from \psr\ relative to the 3.14~s nominal ACIS CCD frametime results
in a $\sim 6\%$ pile-up fraction for the pulsar. This produces evident
distortion in its spectrum that is mitigated by including the
\chandra\ {\tt pileup} model in all spectral fits with ACIS of the
pulsar presented herein.  We compare our joint fits results to those
obtained using data from earlier epochs
\citep{hel03,mor07,kum08,ng09,su09,tem12} as presented in
Table~\ref{tab:psrspec}.


\subsection{Timing Analysis\label{sec:timing}}

In the following timing analysis, all photon arrival times are
converted to barycentric dynamical time (TDB) using the DE405 solar
system ephemeris and the \chandra\ coordinates given in \cite{hel03}.
We extracted photons using an aperture of radius $r<0\farcm75$ and
$r<1\farcm3$ for the \xmm\ and \nustar\ data, respectively, found to
maximize the embedded pulse signal. For \nustar, we include photons
over the full energy band and but restricted our analysis to the
0.5$-$10~keV band for the \xmm\ data to minimize particle
contamination.

\begin{figure}[t]
\vspace{0.1in}
\hfill
\centerline{
\hfill
\includegraphics[angle=270,width=0.9\linewidth]{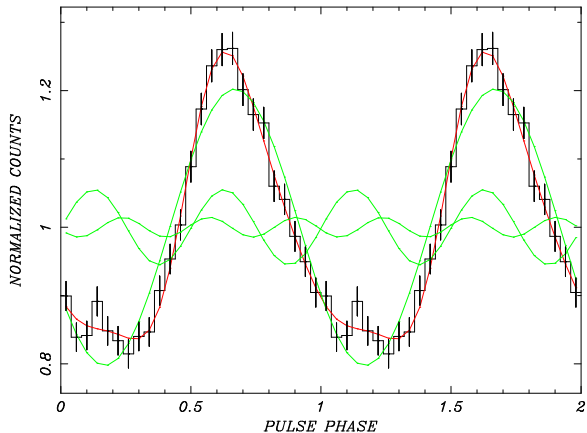}
\hfill
}
\caption{
The background subtracted pulsed profiles of \psr\ using data obtained from the 
\nustar\ observation presented here. Overplotted on the data is a 
model profile (red) composed of the sum of the first three Fourier components (green) 
for the light curve. This profile, for the full energy band, is normalized to 
the average counts per bin.
}
\label{fig:psrfold}
\end{figure}

We searched for the known signal from \psr\ using an accelerated FFT
to account for the substantial frequency derivative associated with
the pulsar. Photon arrival times from the long(er) \nustar\ data were
binned into a 0.01 s light curve and searched at the full resolution
allowed by the span of the data. A highly significant signal is
recovered near the expected frequency and its derivative. We refine
the strongest signal using the $Z_1^2$ test \citep{buc83}, appropriate
for the nearly sinusoidal pulse profile (see
Figure~\ref{fig:psrfold}). In the following phase-resolved
spectroscopy, we use the best-fit period $P=0.32852377(4)$~s and
period derivative $\dot P = 7.4(4)\times10^{-12}$ for epoch MJD~58013.

We also searched from \psr\ for possible millisecond bursts similar to
those detected by \xte\ \citep{gav08}, we examined its light curves
over a range of timescales down to the 10~ms.  We compared the
frequency of occurrence of counts in each light curve bin to those
predicted by Poisson statistics based on the mean count rate and find
no significant outliers.  We therefore conclude that, during the
duration of our \nustar/\xmm\ observation, the pulsar did not display
significant temporal characteristics of a magnetar.

\subsection{Spectral Analysis\label{sec:spectra}}

For spectral fitting we use the {\tt XSPEC} (12.10.0c) package
\citep{Arnaud96} and characterize the column density with the built-in
{\tt TBabs} absorption model, selecting the {\tt wilm} Solar
abundances \citep{Wilms2000} and the {\tt vern} photoionization
cross-section \citep{Verner96}. The $\chi^2$ statistic is used to
evaluate the spectral fits throughout and the parameter uncertainties
are quoted at the 90\% confidence level (C.L.) for one or more
interesting parameters, as appropriate.  Response matrices and
ancillary response files were generated for each data set following
the mission specific standard procedures.  We note that previous
spectral fits in the literature used the {\tt Wabs} absorption model,
a photo-electric absorption using the Wisconsin cross-sections
\citep{mor83}.  Spectral fits using the newer column density model can
result in a measured value that is significantly different, for the
same intrinsic spectrum model.

\subsubsection{\psr\ Pulsed Spectrum}

To isolate the broad-band pulsed component of \psr\ in the \xmm\ and
\nustar\ spectra, we divided the pulse profile into two phase
intervals, corresponding to the peak (``on'' pulse) and the valley
(``off'' pulse) regions. Although the valley spectrum provides a
perfect representation of the unpulsed background to subtracted from
the peak spectrum, it necessarily includes a portion of the pulsed
emission itself. For a theoretical sinusoidal signal, the flux
correction factor for the background subtracted pulsed spectrum is
$\pi/2 = 1.57$. In our case, where the pulse profile is more peaked
than a sine curve, the computed correction factors are 1.41 and 1.30
in the 2$-$10~keV band, for the \xmm\ and \nustar\ data, respectively.
This assumes that the pulse shape is essentially energy independent,
for which there is no significant evidence to the contrary.

We choose source apertures of $r<30^{\prime\prime}$ radius to extract
the pulsar spectra for each mission to help estimate the total flux
from the pulsar and to generate point-source response matrices.  As
the source (``on'') and background (``off'') spectra contain a similar
number of counts per channel, their subtraction results in large
uncertainties. With this in mind, we group all source spectra into
spectral fitting channels that contain a minimum signal-to-noise of 3
sigma after background subtraction.

The joint \xmm\ and \nustar\ pulsed spectrum is well-fit to a simple
absorbed power-law model in the 2$-$10~keV and 3$-$50~keV range,
respectively. The column density is poorly constrained in these fits
and is fixed to $N_{\rm H}=6\times10^{22}$~cm$^{-2}$, the iterated
value obtained from the final, high statistic PWN analysis (described
below). The best-fit photon index is $\Gamma = 1.24\pm0.09$, with a
$\chi^2_{\nu} = 0.992$ for 36 DoF.  The corrected 2$-$10~keV
unabsorbed flux is $F_x = 2.16^{+0.25}_{-0.26}\times
10^{-12}$~erg~s$^{-1}$~cm$^{-2}$ and $F_x = 2.42^{+0.29}_{-0.18}\times
10^{-12}$~erg~s$^{-1}$~cm$^{-2}$ for the \xmm\ and \nustar\ spectral
fits, respectively. These fluxes are self-consistent within the
measurement uncertainties.  We note, moreover, that the mean pulsed
flux of $F_x = (2.25\pm0.35)\times 10^{-12}$~erg~s$^{-1}$~cm$^{-2}$,
is fully consistent with the 2$-$10 keV flux predicted by the curved
power-law model of Kuiper (2018), $F_x = 2.35\times
10^{-12}$~erg~s$^{-1}$~cm$^{-2}$, used to characterize the X-ray to gamma-ray flux.

To estimate the pulsed fraction for \psr, we extracted a total pulsar
spectrum from the 2016 \chandra\ observation using a
$1.5^{\prime\prime}$ aperture and an
$2^{\prime\prime}<r<3^{\prime\prime}$ annular background region. A fit
to an absorbed power-law in the 1$-$10~keV range with the column density fixed at $N_{\rm
  H}=4.0$ yields a best-fit $\Gamma = 1.32\pm0.15$,
with a $\chi^2_{\nu} = 1.19$ for 66 DoF. The total pulsar unabsorbed
flux is $F = (3.1\pm0.1)\times 10^{-12}$~erg~s$^{-1}$~cm$^{-2}$,
implying an intrinsic pulsed fraction of $71\pm16$\% in 2$-$10~keV band,
assuming that the pulsar flux has been steady between the two
observations, which is evidently the case, as shown in the next
section.
 
To compare our pulsed measurements to published results on \psr\
obtained at earlier epochs we re-fitting these spectra using the
historic {\tt XSPEC} {\tt Wabs} model for the interstellar absorption
\cite{hel03,kum08,ng08}.  These results are summarized in
Table~\ref{tab:psrspec}.

\begin{figure}[t]
\hfill
\centerline{
\hfill
\includegraphics[angle=270,width=0.98\linewidth]{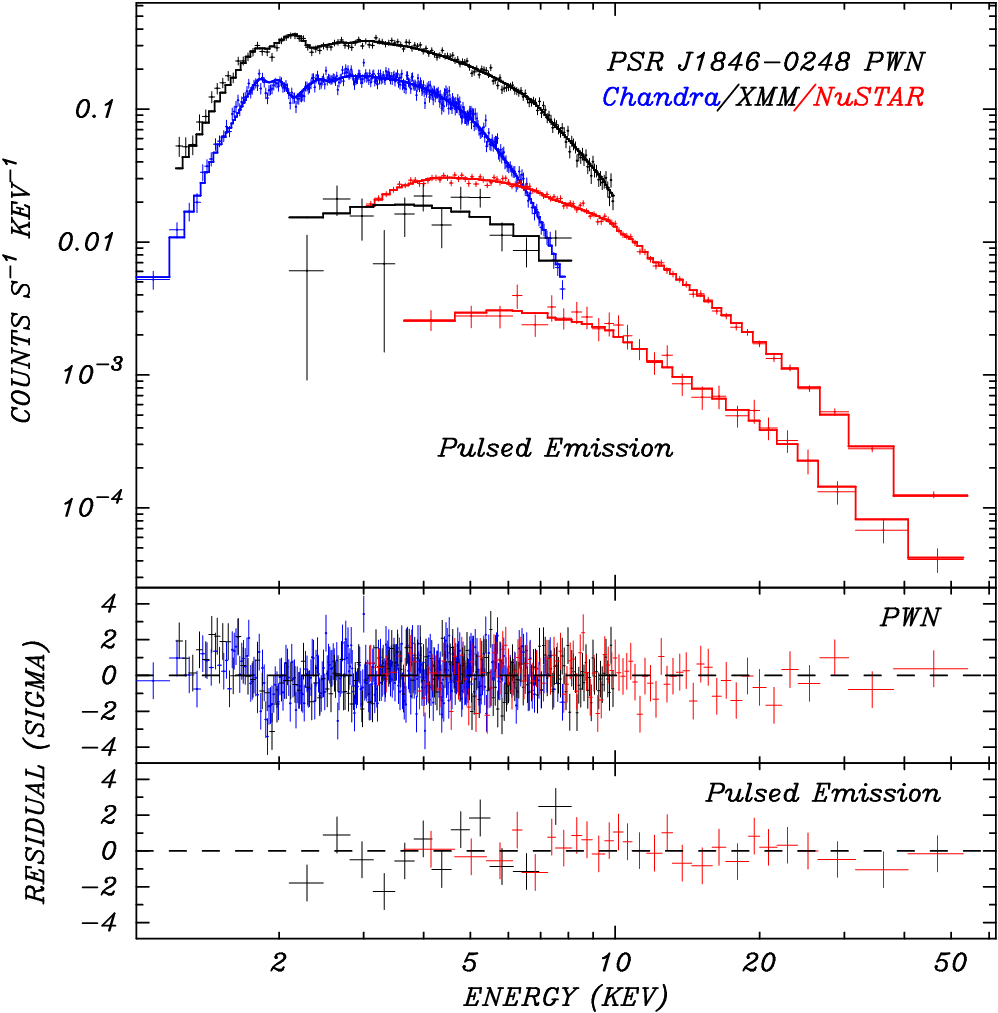}
\hfill
}
\caption{ Broad-band X-ray spectra of \psr\ and its PWN in SNR \snr.
  The \chandra\ (blue), and concurrent \xmm\ EPIC~pn (black) and
  \nustar\ (red) Kes~75 PWN spectra are fitted jointly (top three
  curves). Similarly for the \xmm\ EPIC~pn and \nustar\ phase-resolved
  pulsed spectra (lower two curves). Both sets of spectra are fitted
  to an absorbed power-law model with independent normalizations.  For
  the PWN spectra, the \xmm\ model included a component to account for
  the pulsar emission below $\sim$2~keV and the \nustar\ spectrum
  includes a component for the pulsed emission, significant $>30$~keV
  (see text for details).  Upper Panel --- the data points ({\it
    crosses}) are plotted along with the best fit model ({\it
    histogram}) given in Table~\ref{tab:pwnspec}.  Lower panels ---
  the best fit residuals for PWN (upper) and pulsed emission (lower)
  spectra in units of sigma.  }
\label{fig:spectra}
\end{figure}

\begin{deluxetable}{lccc}[!ht]
\footnotesize
\tablewidth{0pt}
\tablecolumns{4} 
\tablecaption{Spectra of \psr\ in \snr}
\tablehead{
\colhead{Reference} & \colhead{$N_{\rm H}$}          & \colhead{$\Gamma$} & \colhead{Flux$^{a}$}  \\
                    & \colhead{($10^{22}$ cm$^{-2}$)}&                    & \colhead{[$\times10^{-12}$]}
}
\startdata\\[-0.65em] 
\multicolumn{4}{c}{\chandra\ 2000 data set}\\[0.35em]
\cline{1-1} \cline{2-2} \cline{3-3} \cline{4-4} \\[-0.65em]
Helfand \etal\ 2003  & 3.96 (fixed)        & $1.39\pm0.04$        & $7.1$            \\
Ng \etal\ 2008       & 4.0 (fixed)         & $1.1\pm0.01$         & $6.1\pm0.03$     \\
Gavriil \etal\ 2008  & \dots               & $1.17_{-0.12}^{+0.15}$   & \dots            \\
Kumar \etal\ 2008    & 3.96 (fixed)        & $1.32_{-0.09}^{+0.08}$   & $4.3\pm0.2$      \\
\cutinhead{\chandra\ 2006 data set (Flare Epoch)}
Ng \etal\ 2008       &  4.0 (fixed)        & $1.86\pm0.02$        & $37\pm01.0$     \\
Gavriil \etal\ 2008&  \dots                & $1.89_{-0.06}^{+0.04}$   & \dots            \\
Kumar \etal\ 2008    &  $4.15_{-0.12}^{+0.09}$ & $1.97_{-0.07}^{+0.05}$   & $27_{-2}^{+1}$  \\
\cutinhead{\chandra\ 2016 data set}
This Work            &  4.0 (fixed)       &  $1.32\pm0.15$       & $3.1\pm0.1$   \\
\cutinhead{\xmm\ \& \nustar\ 2017 data sets (Pulsed Only) }
This Work           & 4.0 (fixed)         & $1.23\pm0.09$        & $2.2\pm0.4$  \\
\enddata 
\label{tab:psrspec}
\tablenotetext{}{\footnotesize Column density is derived using the built-in 
{\tt XSPEC} {\tt Wabs} interstellar absorption  model in all cases for comparison purpose. 
Quoted uncertainties for 90\% C.L. for one interesting parameter.}
\tablenotetext{a}{\footnotesize Unabsorbed 2$-$10 keV flux, in units of erg cm$^{-2}$ s$^{-1}$.}
\end{deluxetable}

\subsubsection{\psr\ PWN Spectrum}

As noted above, it is not possible to spatially isolate the PWN from
the pulsar emission in the \xmm\ and \nustar\ data sets. Instead, we
fit the composite spectrum and account for the pulsar emission in the
source aperture with an additional power-law component. For \xmm, we
estimate the pulsar contribution using the spatial resolved \chandra\
results, and for \nustar, we use the measured pulsed spectrum, both
presented above. This is necessary to account for significant
contaminated below $\sim$2~keV in the \xmm\ PWN spectrum from the
pulsar and its environment, and for the pulsed emission that distorts
the \nustar\ spectrum above $\sim$30~keV.  For the \xmm\ spectra we
use a $r<30^{\prime\prime}$ source aperture that captures most of the
PWN flux and averages over its radial-dependence spectrum. For the
\nustar\ spectrum, with its poorer spatial resolution, we use a
smaller $r<24^{\prime\prime}$ source aperture, to account for the
additional blur of the PWN flux in order to better match the \xmm\
spectrum extraction region, while simultaneously minimizing possible
contamination from the nearby SNR lobes. For both missions, we
estimate the SNR background in the source aperture using a concentric
annular background region $33^{\prime\prime}< r <45^{\prime\prime}$.
We note that background contribution is small ($<$10\%) except below
$\sim$2~keV. As the PWN is poorly resolved spatially in both
telescopes, we again use point-source mirror response matrices to
characterize the effective area.

The spatially averaged \xmm\ and \nustar\ PWN spectra are fitted
jointly with their normalization left free to allow for systematic
differences in the aperture flux. The spectrum is well-modelled by an
absorbed power-law over the 1.2$-$50~keV span of the two data sets.
This yields a column density $N_H = (6.0\pm0.1) \times
10^{22}$~cm$^{-2}$ and photon index $\Gamma = 2.03\pm0.02$, with a
$\chi^2_{\nu} = 1.04$ for 399 degrees-of-freedom (DoF). We note that,
without taking into account the pulsar component in these spectra it
is not possible to get a consistent power-law index in the overlapping
3$-$10~keV band.  To check for evidence of spectral curvature we
fitted a broken power-law model to the joint spectrum, however, no
significant change in the index is measurable in the current data.

We next compare the 2016 \chandra\ PWN spectrum to the above \xmm\
result, over their mutual energy range. A spectra was extracted from
the $r<30^{\prime\prime}$ \xmm\ region, excluding a
$r<2^{\prime\prime}$ circle around the pulsar. The best fit power-law
model in the 1.0--8~keV band yields $N_H = (5.9\pm0.2) \times
10^{22}$~cm$^{-2}$ and photon index $\Gamma = 2.00\pm0.06$, with a
$\chi^2_{\nu} = 1.07$ for 306 DoF.  These parameters are consistent
with the \xmm\ results, suggesting no spectral change between the two
epochs, and if so, the \chandra\ flux is considered the most accurate.
It is reassuring that the details of the residuals of both spectra are
effectively identical. The common systematic deviations in the spectra
suggest that they are related to differences in the SNR emission
between the source and background regions.

Finally, to obtain the most accurate PWN spectrum and flux
measurement, we re-fitted the joint \xmm\ and \nustar\ PWN spectra
with the addition of the \chandra\ spectrum, again with independent
normalizations. The best-fit parameters are $N_H = (6.0\pm0.1) \times
10^{22}$~cm$^{-2}$ and photon index $\Gamma = 2.0\pm0.02$, with a
$\chi^2_{\nu} = 1.075$ for 548 DoF. The final 2--10 keV unabsorbed PWN
flux of $(2.13\pm0.03)\times 10^{-11}$~erg~s$^{-1}$~cm$^{-2}$, is
determined from the \chandra\ component of this fit, after allowing
for the missing flux from the excluded pulsar region, a $\sim$10\%
effect. This missing flux is estimated from a spectrum of the bright
northern PWN knot, extracted from a $r<2^{\prime\prime}$ aperture. The
ratio of the pulsar to the total flux (pulsar+PWN) is consistent with
the observed pulsed fraction $\sim$10\% in the 2$-$10~keV band.  The
pulsar and PWN spectral results are reported in
Table~\ref{tab:pwnspec}

\begin{figure*}[t]
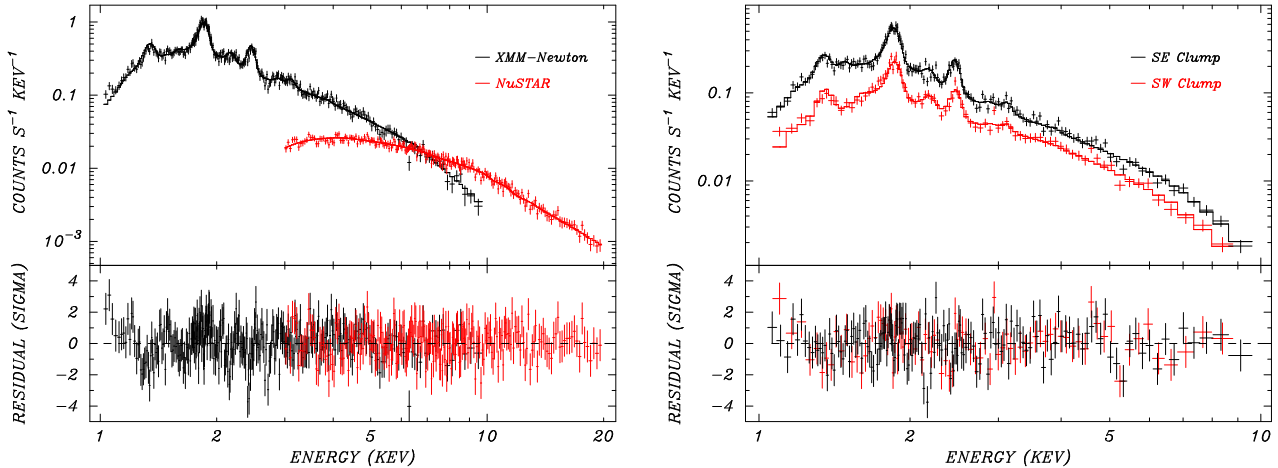

\vspace{0.1in}
\hfill
\centerline{
\hfill
\includegraphics[angle=270,width=0.45\linewidth]{./kes75_snr_pn_fpm_spec.eps}
\hfill
\includegraphics[angle=270,width=0.45\linewidth]{./kes75_snr_clumps_pn.eps}
\hfill }
\caption{ The spectrum of \snr\ fitted to an absorbed non-equilibrium
  thermal plasma ({\tt NEI}) plus power-law model. Right --- Joint fit
  to the concurrent \xmm\ EPIC pn ({\it black}) and \nustar\ ({\it
    red}) spectra with independent normalizations. The EPIC~pn data
  that covers the bulk of the bright sector, as described in the text.
  Left --- The \xmm\ EPIC~pn X-ray spectrum of the bright thermal
  emission from the southeast clump ({\it black}) and the { southwest}
  clump ({\it red}). Upper Panels --- data points ({\it crosses})
  plotted along with the best fit models ({\it histogram}) are given
  in Table~\ref{tab:snrspecrev}.  Lower panels --- the best fit
  residuals in units of sigma.  }
 \end{figure*}

\begin{deluxetable}{lccc}[!ht]
\small
\tablewidth{0pt}
\tablecaption{Spectra of the pulsar and PWN in \snr}
\tablehead{
\colhead{Parameter} & \multicolumn{2}{c}{Pulsar}  & \colhead{PWN}\\
                    & \colhead{Pulsed} & \colhead{Total}  & \colhead{} 
}
\startdata
$N_{\rm H}$~($10^{22}$ cm$^{-2}$) & $6.0$ (fixed)      & $6.0$(fixed)      & $6.0\pm0.1$        \\ 
Photon Index $\Gamma$             & $1.24\pm0.09$      & $1.38\pm0.16$      & $2.03\pm0.02$        \\
$F[\times10^{-11}]^{a}$ (abs.)    & $0.18\pm0.03$      & $0.24\pm0.01$      & $1.53\pm0.02$        \\
$F[\times10^{-11}]^{b}$ (unabs.)  & $0.23\pm0.04$      & $0.31\pm0.01$    & $2.13\pm0.02$        \\
Luminosity $L_{\rm x}^{c}$        & $9.9\times10^{33}$ & $1.3\times10^{34}$  & $9.2\times10^{34}$   \\ 
$\chi^2_{\nu}(DoF)$               & 0.99(36)           & 1.19(66)         & 1.04(399)         \\
\enddata
\label{tab:pwnspec}
\tablenotetext{}{\footnotesize Note --- Quoted uncertainties for 90\%
  C.L. for one or two interesting parameters, for the pulsar and PWN,
  respectively. Pulsed emission is from the a joint \xmm/\nustar\
  phase-resolved spectroscopy; the total pulsar flux is determined
  from a fit to the \chandra\ data (see text for details)}
\tablenotetext{a}{\footnotesize Absorbed 2$-$10 keV flux, in erg  cm$^{-2}$ s$^{-1}$.}  
\tablenotetext{b}{\footnotesize Unabsorbed  2$-$10 keV flux, in erg~s$^{-1}$.}  
\tablenotetext{c}{\footnotesize 2$-$10 luminosity, in erg~s$^{-1}$, for $d = 6$~kpc.}
\end{deluxetable}

\subsubsection{\snr\ SNR Spectrum \label{sec:snr}}

We also extracted \xmm\ and \nustar\ spectra for the SNR emission
using an $0\farcm9<r<1\farcm8$ source annulus and
$2^{\prime}<r<3^{\prime}$ background region (see
Figure~\ref{fig:images}). This covers the two bright X-ray clumps
labeled southeast (SE) clump and southwest (SW) clump in \cite{hel03},
encompassing the bulk of the remnant emission.  For \xmm, in small
window mode, the SNR fell partially off the field of view but still
includes the clumps.

\begin{deluxetable}{lccc}[!ht]
\tablewidth{0pt}
\tablecaption{Spatially resolved spectra of SNR \snr}
\tablehead{
\colhead{}                        & \colhead{Southeast} & \colhead{Southwest} & \colhead{Sector$^{a}$}
}
\startdata 
Region R.A.                & 18:46:22.764                 & 18:46:28.574               &18:46:24.893 \\
Region Decl.               & $-$02:59:43.33               & $-$02:59:09.53             & $-$02:58:28.31 \\
Ellipse radii              &  $1\farcm0\times0\farcm4$    & $0\farcm8\times 0\farcm45$ & \dots \\ 
Ellipse P.A.               &  $300^{\circ}$               &  $40^{\circ}$              & \dots \\ 
Annulus radii              &  \dots                       & \dots                      & $0\farcm9\times 1\farcm8$ \\
\tableline
$N_{\rm H}$~$(10^{22}$~cm$^{-2}$) & $4.11^{+0.11}_{-0.10}$ &  $4.27^{+0.23}_{-0.31}$
& $4.14^{+0.11}_{-0.09}$ \\
$kT~(keV)$          & $1.01^{+0.17}_{-0.10}$ & $1.18^{+0.24}_{-0.20}$
& $0.87^{+0.05}_{-0.06}$ \\
$\tau$~($10^{10}$~s~cm$^{-3}$)    & 8.5 & 4.2 
& $10.0^{+2.2}_{-1.6}$ \\
Mg   & $0.89^{+0.09}_{-0.11}$  & $0.76^{+0.20}_{-0.16}$ & $0.81\pm0.08$ \\
Si   & $1.26^{+0.15}_{-0.13}$  & $1.21^{+0.46}_{-0.28}$ & $1.27\pm0.09$ \\
S    & $0.98^{+0.17}_{-0.14}$  &  $1.0^{+0.53}_{-0.27}$ & $1.15\pm0.11$ \\
NEI Flux ($10^{-11})^{b}$  & 3.2  & 1.7                &  10 \\
Photon Index $\Gamma$          &  $1.73^{+0.33}_{-0.63}$&      $2.02^{+0.37}_{-0.92}$  
& $2.02^{+0.04}_{-0.05}$ \\
PL Flux ($10^{-12})^{b}$  & 1.6  & 1.3                &  $8.7$ \\
$\chi^2_{\nu}(DoF)$    &   1.459 (452) &  1.369 (363)
& 1.369 (896) \\ 
\enddata
\label{tab:snrspecrev}
\tablenotetext{}{\footnotesize See Figure~\ref{fig:images} for the SNR clump regions. Coordinates are in the J2000 system. Quoted uncertainties for 90\% C.L. for the parameter of interest.}
\tablenotetext{a}{\footnotesize  Joint fit to \xmm\ EPIC~pn, \chandra\ and \nustar\ spectra, with independent normalizations.}
\tablenotetext{b}{\footnotesize  Unabsorbed \xmm\ EPIC pn flux measured in the 0.5$-$10 keV band in units of erg~s$^{-1}$~cm$^{-2}$. The first two columns correspond to the clumps fits using the \xmm\ and \chandra\ dataset.}
\end{deluxetable}

The spectra of the SNR were fitted to a non-equilibrium thermal plasma
model ({\tt XSPEC NEI}) with variable abundances plus a power-law
component (see Table~3), as indicated in the first \chandra\ study by
\cite{hel03}.  This model produces an excellent fit, but with
parameter values notably different from those reported in the earlier
\chandra\ studies \citep[e.g.,][]{hel03,tem12}.  This can be
attributed, at least in part, to the updated ISM model used herein and
to the difference in the extraction regions.

We next included the \chandra\ data described in \S2.3 to the joint
\xmm\ and \nustar\ fit, to verify \xmm\ spectrum in their common
energy range and for an improved flux measurement.  We find that that:
a) a single component ({\tt XSPEC NEI}) with variable abundances does
model the data well, confirming the need for an additional harder
component; b) both a two-temperature {\tt NEI} model and an {\tt
  NEI}+power-law model provide an improved fit to the spectra.
However, the \xmm's greater sensitivity to the hard diffuse X-ray
emission, along with the expanded \nustar\ energy band allowed us to
clearly select the power-law model as prefer over the thermal NEI
model.  The best fit model parameters obtained with the joint fit are
presented in Table~\ref{tab:snrspecrev}.

As noted above, the plasma temperature of the NEI model (softer
component) differs because of the chosen {\tt TBabs} with the {\tt
  wilm} abundances. Previous works used the \cite{angre89} abundances
accounting for the lower column density and lower temperatures.  We
confirm this by checking against the results obtained using the early
\chandra\ observation ({\tt Obsid 748}), reprocesses with the latest
calibrations.

\begin{figure*}[t]
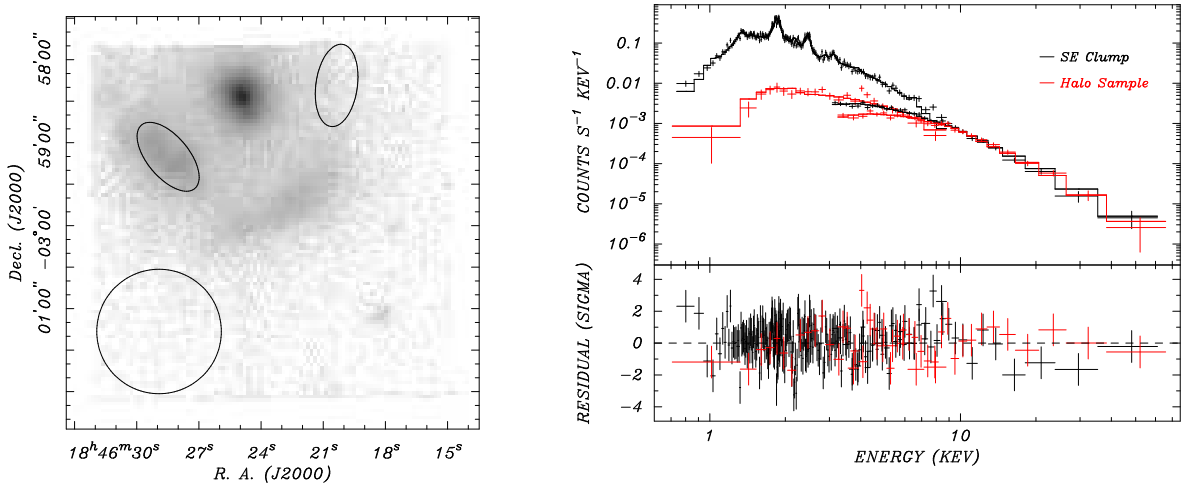

\hfill
\centerline{
\hfill
\includegraphics[angle=270,width=0.35\linewidth]{./kes75_halo_xmm_epn_1000_10000_img_corr_fixed.eps}
\hfill
\includegraphics[angle=270,width=0.45\linewidth]{./kes75_lobe_halo_comparison.eps}
\hfill
}
\caption{Evidence of a dust scattered PWN halo component in hard SNR spectra of \snr. 
{Left} --- An \xmm\ image of the \snr\ SNR showing the location of mirrored source regions, situated on opposite sides of the SNR, one centered on the SE clump (left) and the other placed to sample the halo region (right), equi-distant from the pulsar. The large circle shows the background region.  {Right}  --- A joint fit to the concurrent \xmm\ and \nustar\ spectra extracted from the two regions fitted with the {\tt NEI} plus  power-law model with their power-law indices linked. There is no evidence for excess thermal emission in the halo region and the flux from this region is sufficient to account for the hard continuum component in the SE Clump spectrum.
}
\label{fig:halo}
\end{figure*}

The background for each spectrum is generally less than an order of
magnitude of the source flux, up to about 20~keV, beyond which it
dominates the \nustar\ spectrum. Small systematic errors in the
power-law index are possible due to different in the size of the PSF
for the two missions over the source aperture.  However, this is
likely to be a small effect because emission from outside the source
aperture still contributes proportionally more by area to offset the
losses from the larger surface brightness emission inside the
aperture.

Using the \xmm\ data alone, we also analysed the two clump regions
separately.  Because of its larger PSF, the clumps are not well
isolated in the \nustar\ data, and we do not attempt a joint spectral
fit with the \xmm\ data.  The clump extraction regions and results are
presented in Table~\ref{tab:snrspecrev}.  An annular
($2\farcm2<r<3\farcm0$) background region, centered at coordinates
18:46:24.862~$-$02:58:28.312 (J2000), is used for both clumps.

We have also attempted a two-temperature component fit as done in
previous studies (e.g. Temim \etal\ 2012). We find that the spectra
require temperatures of $\sim$0.9~keV and 4.6~keV for the cool and
hotter component, respectively, of the southeast clump, and
$\sim$0.85~keV and 3.2~keV for southwest clump. However, the reduced
chi-squared in the 2$T$-component fits is higher ($\geq$1.5) than for
the thermal+power-law model, and the {\tt vnei} ionization timescale
is too low in the 2$T$-component fits. We therefore again favor the
power-law interpretation of the hard component. This result is
consistent with the joint fit of the SNR sector with the \nustar\
data.  Below, and in \S4.2, we discuss the nature of the hard
component and link it to a dust scattered PWN halo.

\subsection{Evidence for a Dust Scattered PWN Halo}

The need for a highly significant hard continuum component to model
the SNR spectra of \snr\ has been evident since the first spatially
resolved X-ray observations reported by \cite{hel03}, and later by
\cite{su08,tem12}.  These authors considered several possible origins
for this non-thermal emission and concluded that a dust scattered PWN
halo is the most consistent with the data. Furthermore, by comparing
\chandra\ observations taken during and post flare epochs,
\cite{reynolds18} found spatial evidence for a "transient halo",
likely depended on the brightness of the pulsar/PWN.

To better understand this hard component we compare broad-band spectra
of the SNR's SE clump with that obtained from a mirrored region on the
other side of the SNR, that lacks evidence of thermal emission, to
represent the putative halo emission. For the SE SNR clump we
extracted spectra from a $15^{\prime\prime}\times30^{\prime\prime}$
region, smaller then used in Section~\ref{sec:snr}, to minimize the
background contribution to the source region. As shown in
Figure~\ref{fig:halo}, the halo and the SE clump regions are
equi-distant from the pulsar; a background region is chosen well away
from the SNR, to allow for the instrumental signature and the local
Galactic ridge emission.

Figure~\ref{fig:halo} presents the joint fit to the \xmm\ and \nustar\
spectra using the NEI plus power-law model, with the photon indices
for the clump and the halo spectra linked; the thermal component is
set to zero for the halo fits.  The \nustar\ data is crucial to
isolate the higher-energy component in a band where it is most
dominant.  As expected, this fit to the SE clump spectrum reproduced
that reported in Table~\ref{tab:snrspecrev}. Notably, the best-fit
model yields a power-law index of $\Gamma = 1.97\pm 0.09$, consistent
with that found for the PWN spectrum. The averaged 2--10~keV surface
brightness in the halo aperture $\sim$$1\farcm2$ away from the pulsar
is $6.0\times10^{29}$~erg/s/arcsec$^2$ at 6~kpc. Most importantly, the
flux from the clump and the halo spectrum in the \nustar\ band is
essentially the same (see Figure~\ref{fig:halo}).

These results suggest that most, if not all, of the hard emission
component for the SNR clumps can be attributed to a non-localized
spectral component.  To consider the radial dependence of this
component, we examined a \chandra\ spectrum obtained from the
$33^{\prime\prime}< r <45^{\prime\prime}$ annular region between the
PWN and the clumps. This region is found to be an add-mixture of the
thermal component and the power-law emission, with a similarly hard
photon index ($\sim$2), and a surface brightness that is approximately
a factor of 3 times larger than that of the western halo region. The
decrease in surface brightness of the hard component away from the
pulsar is consistent with a dust-scattered PWN halo interpretation, as
further discussed in section~\ref{sec:discussion}.

\section{PWN-SNR Modeling\label{sec:modeling}}

\begin{figure}[tb]
  \begin{center}
    \includegraphics[width=0.475\textwidth]{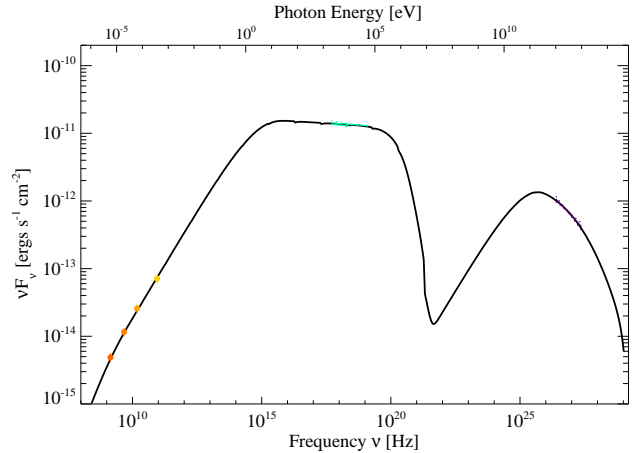}
  \end{center}
  \vspace*{-0.25cm}
  \caption{Observed (color) and predicted (black line) spectral energy
  distribution (SED) of the PWN in Kes 75.  The observed and predicted
  quantities are provided in Table \ref{tab:obsprop}, while the model
  parameters are shown in Table \ref{tab:modprop}.  }
  \label{fig:bestspec}
\end{figure}

Currently, the best way of determining the energetics of the \snr\
progenitor, the birth parameters of \psr, and for its pulsar wind, is
to fit the observed properties of the PWN with a model for its
dynamical and radiative evolution (see \citealt{gelfand17} for a
recent review of such models).  Following our previous analysis of
\snr\ \citep{gelfand14} and similar SNR/PWN systems (e.g., G54.1+0.3,
\citealt{gelfand15}; HESS~J1640$-$465, \citealt{gotthelf14}), we use a
Markoff Chain Monte Carlo code to determine the combination of input
parameters of such a model (based on that described by
\citealt{gelfand09}) best reproduce (lowest $\chi^2$) the observed
properties of Kes~75 listed in Table \ref{tab:obsprop}.  The input
parameters are listed in Table \ref{tab:modprop}, with the age $t_{\rm
  age}$ and initial spin-down luminosity $\dot{E}_0$ chosen that, for
a given braking index $p$ and spin-down timescale $\tau_{\rm sd}$,
reproduce the pulsar's current characteristic age $t_{\rm ch}$ and
spin-down luminosity $\dot{E}$:
\begin{eqnarray}
  \label{eqn:tage}
  t_{\rm age} & = & \frac{2t_{\rm ch}}{p-1}-\tau_{\rm sd} \\
  \label{eqn:e0dot}
  \dot{E}_0 & = &  \dot{E} \left(1+\frac{t_{\rm age}}{\tau_{\rm
      sd}}\right)^{\frac{p+1}{p-1}},
\end{eqnarray}
the PWN's inverse Compton emission is the result of leptons scattering
of the Cosmic Microwave Background and an additional photon field with
temperature $T_{\rm ic}$ and normalization $K_{\rm ic}$, defined such
that the energy density $u_{\rm ic}$ of this photon field is:
\begin{eqnarray}
  \label{eqn:uic}
  u_{\rm ic} & = & K_{\rm ic} a_{\rm bb} T_{\rm ic}^4,
\end{eqnarray}
where $a_{\rm bb} = 7.5657\times10^{-15}~\frac{\rm erg}{\rm
  cm^3~K^4}$, as well as assuming the spectrum of particle injected
into the PWN at the termination is well described by a broken
power-law:
\begin{eqnarray}
  \label{eqn:injspec}
  \frac{d\dot{N}_{e^\pm}}{dE}(E) & = & \left\{
  \begin{array}{cc} \dot{N}_{\rm break} \left(\frac{E}{E_{\rm break}} \right)^{-p_1}
    & E_{\rm min} < E < E_{\rm break} \\
    \dot{N}_{\rm break} \left(\frac{E}{E_{\rm break}} \right)^{-p_2} &
    E_{\rm break} < E < E_{\rm max}
  \end{array}
  \right. , \\
\end{eqnarray}
where $\dot{N}_{e^\pm}$ is the rate $e^\pm$ are injected into the PWN
and $\dot{N}_{\rm break}$ is calculated requiring that at all times:
\begin{eqnarray}
\label{eqn:ndotbreak}
(1-\eta_{\rm B})\dot{E} & \equiv & \int\limits_{E_{\rm min}}^{E_{\rm
    max}} E \frac{d\dot{N}_{e^\pm}}{dE} dE,
\end{eqnarray}
where the magnetization of the wind $\eta_{\rm B}$ is defined to be
the fraction of the pulsar's spin-down luminosity injected into the
PWN as magnetic fields.  Unfortunately, the number of model parameters
is one more than the number of observed quantities for this system 
, and the set of input parameters
which produced the lowest $\chi^2$ ($\chi^2 = 0.93$) are provided in 
Table \ref{tab:modprop} with the observed and predicted spectral
energy distribution (SED) of this source shown in Figure
\ref{fig:bestspec}. The significance and limitations of these results are discussed below.

\begin{deluxetable}{cccc}[!ht] 
\tablewidth{0pt}
\tablecolumns{4} 
\tablecaption{Observed and Predicted PWN-SNR Model for Kes~75}
\tablehead{
\colhead{Property}              & \colhead{Observed}    & \colhead{Model}  & \colhead{Reference}
}
\startdata \\[-0.65em] 
\multicolumn{4}{c}{SNR}\\[0.35em]
\cline{1-1} \cline{2-2} \cline{3-3} \cline{4-4} \\[-0.65em] 
      Radius ($^{\prime}$)      & $1.5\pm0.15$          & $1.50$            & $\cdots$ \\ 
      Distance (kpc)            & $5.8_{-0.4}^{+0.5}$   & $5.8$            & \citealt{verbiest12} \\
 \cutinhead{PWN}
      Radius  ($^{\prime\prime}$)& $30\pm1.67$           & $30.0$           & $\cdots$ \\
      $\dot{\theta}_{\rm pwn}$ (\%/yr) & $0.249\pm0.023$ & $0.234$          & \citealt{reynolds18} \\ 
      $S_{1.4}$ (mJy)           & $348\pm52$             & 327              & \citealt{salter89} \\
      $S_{4.7}$ (mJy)           & $247\pm37$             & 240              & \citealt{salter89} \\
      $S_{15}$ (mJy)            & $172\pm26$             & 158              & \citealt{salter89} \\
      $S_{89}$ (mJy)            & $80\pm12$              & 82               & \citealt{bock05}\\
      $\Gamma$(2$-$55~keV)      & $2.031\pm0.025$          & $2.030$           & $\cdots$ \\
      $F$(2$-$10~keV)$^a$        & $2.13\pm0.022$        & $2.133$          & $\cdots$ \\ 
      $\Gamma$(1$-$10~TeV)      & $2.41\pm0.01$          & 2.42             & \citealt{hess18}      \\
      $F$(1$-$10~TeV)$^a$       & $0.160\pm0.016$        & $0.159$          & \citealt{hess18} \\
 \cutinhead{PSR}
      $\dot{E}$ (erg~s$^{-1}$)  & $8.10\times10^{36}$    & $\cdots$         &  \citealt{livingstone11} \\
      $t_{\rm ch}$ (yr)         & 728                    & $\cdots$         & \citealt{livingstone11} \\
      $p$                       & $2.65\pm0.01$          & $2.652$          & \citealt{livingstone11} \\
\enddata
\tablenotetext{}{\footnotesize  Notes --- The predicted values are for
  the set of model parameters which resulted in the lowest $\chi^2$.
  Quantities derived in this paper have $\cdots$ for their reference.}
\tablenotetext{a}{\footnotesize Flux in units of $10^{-11}$~erg~cm$^{-2}$~s$^{-1}$}
\tablenotetext{b}{\footnotesize {www.atnf.csiro.au/research/pulsar/psrcat} \citep{manchester2005}}

\label{tab:obsprop}
\end{deluxetable}

\begin{deluxetable}{cc}[!ht] 
\tablewidth{0pt}
\tablecolumns{2} 
\tablecaption{Best fit PWN-SNR Model Parameters for \snr}
\tablehead{
      \colhead{ Model Parameter} & \colhead{ Min. $\chi^2$ Value}
}
\startdata
      Supernova Explosion Energy $E_{\rm sn}$            & $1.26\times10^{50}~{\rm erg}$ \\
      Supernova Ejecta Mass $M_{\rm ej}$                 & $0.51~{\rm M}_\odot$ \\
      ISM density $n_{\rm ism}$                          & $0.56~{\rm cm}^{-3}$ \\
      Pulsar braking index $p$                           & 2.652 \\
      Pulsar spin-down timescale $\tau_{\rm sd}$         & 398~yr \\
      Age $t_{\rm age}$                                  & 483~yr \\
      Initial Spin-down Luminosity $\dot{E}_0$           & $4.69\times10^{37}~\frac{\rm erg}{\rm s}$ \\
      Pulsar Wind Magnetization $\eta_{\rm B}$           & 0.0724 \\
      Min. energy of injected particles $E_{\rm min}$    & 2.00~GeV \\
      Break energy of injected particles $E_{\rm break}$ & 2042~GeV \\
      Max. energy of injected particles $E_{\rm max}$    & 1.00~PeV \\
      Low-energy index of injected particles $p_1$       & 1.73 \\
      High-energy index of injected particles $p_2$      & 3.04 \\
      Temperature of IC photon field $T_{\rm ic}$        & 32~K \\
      Normalization of IC photon field $K_{\rm ic}$      & $1.17\times10^{-3}$ \\
\enddata
\tablenotetext{}{\footnotesize  Notes --- Parameters of the model for the evolution of a PWN inside a
  SNR used to reproduced the observed properties of Kes~75 (Table
  \ref{tab:obsprop}), as well as the values for the combination of
  parameters which give the lowest $\chi^2$.}
\label{tab:modprop}
\end{deluxetable}

\section{Discussion and Conclusions\label{sec:discussion}}

Based on the 2016 \chandra\ and 2017 \xmm\ and \nustar\ observations
of \psr\ reported herein, the pulsar had returned to its quiescent
rotation-powered state following the last known, 2006, magnetar-like
event. In this state, we show that there is no evidence in the pulsed
emission for an additional, flatter spectral component above 10~keV,
as reported for many magnetars (\citealt[e.g.,][]{den08}; see reviews
by \citealt{mer15,kas17}). The pulsar is found to be highly modulated
in the 2--10 keV band with a lower limit of at least 64\%, not
atypical for young X-ray pulsars with sinusoidal pulse profiles.  The
new, high quality spectral measurements confirms the relation between
the power-law slopes for \psr\ and its PWN, consistent with that
predicted for other highly energetic rotation-powered pulsars
\citep{got03}.  The flux from the PWN is consistent with that reported
by \cite{reynolds18} and we find no evidence of curvature in its
broad-band 1--55~keV spectrum.

The efficiency of powering the PWN from spin-down radiative losses,
($L_x/\dot E \approx 1\%$), estimated from the 2--10~keV luminosity at
distance of 6~kpc \citep{lea07,kum08}, is among the highest known for
rotation-powered pulsars. The discovery of magnetar activity suggest
that at least part of the pulsar luminosity could be from surface
emission from the pulsar, however, the spectrum of the pulsar in the
quiescent state, quite distinct from that expected from a magnetar,
typically characterize by a hot, 0.5~keV blackbody emission and a
steep non-thermal component $\Gamma \sim$4 \citep{par98,mar01}.  It is
possible that some of the flux, consistent with the pulsed fraction,
could be due to a cooling NS, heated during the magnetar activity.
Unpulsed, soft blackbody emission they could be responsible for the
lower column density measured for the pulsar relative to that of the
PWN, $N_{\rm H} \sim 4 \times 10^{22}$~cm$^{-2}$ and $\sim 6 \times
10^{22}$~cm$^{-2}$, respectively. However, the high column density and
instrument band-pass limits our ability to investigation this further.

The broad-band X-ray spectroscopy of \snr\ SNR reveals for the first
time that the hard continuum component is dominant above 10~keV, and
is clearly independent of the thermal emission.  Similar non-thermal
components has been detected from several SNR shells, as first
discovered in the specially-resolved spectroscopy of SN~1006
\citep{koy95}.  This emission is typically described in the X-ray band
by a power-law with a steeper photon index $\sim$2.5--3 \citep{rey08}
and can result in particle acceleration up to TeV energies
\citep[e.g.,][]{rey11},

For \snr, the flatter spectrum and radial fall-off of the hard
emission supports a dust scattering halo origin for this component,
that can be attributed to the bright and heavily absorbed pulsar/PWN
\citep{hel03,su09,tem12}. Similar halos have been discovered around
magnetars or highly magnetized neutron stars that display a
magnetar-like outburst \citep[e.g.,][]{esp13,saf13}.  In the case of
Kes~75, a brightning of the halo was noted following the 2006 flare
\citep{reynolds18}.  The current work shows that the halo emission is
detectable even during the quiescent state of the pulsar.  Such a halo
has been also seen from G21.5--0.9, a similarly young, bright and
heavily absorbed PWN \citep[e.g.,][]{mat10,boc05}.

Making use of the new X-ray results for \snr, we re-evaluated the
evolution of the PWN in the SNR using the dynamical and radiative
model described in Section~\ref{sec:modeling}.  The lack of degree of
freedoms for this model makes it difficult to draw statistically
meaningful results from the fitting, however, the preferred parameters
provide insight into the origin and underlying physics in this system.
The results of our modeling strongly prefers Kes 75 as originated in a
low energy ($E_{\rm sn} \ll 10^{51}~{\rm ergs}$), low ejecta mass
($M_{\rm ej} < 1~{\rm M}_\odot$) explosion.  These quantities are
determined in part by the observed high expansion rate of the PWN
(Table \ref{tab:obsprop}), which, as discussed by \citet{reynolds18},
implies the PWN is expanding into low density supernova ejecta.  One
possibility, as mentioned by \citet{reynolds18}, is the PWN is
currently embedded in a low density bubble resulting from the decay of
$^{56}$Ni in the innermost ejecta (\citealt{li93, chevalier05}).
However, our modeling also simultaneously reproduces the observed size
of the PWN and SNR, which is less dependent on the local conditions
around the nebula.

Taken at face value, the low values of $E_{\rm sn}$ and $M_{\rm ej}$
have strong implications for its progenitor.  The value of $M_{\rm
  ej}$ is lower than what would be expected for an isolated massive
star progenitor (e.g., \citealt{sukhbold16, raithel18}), even when
considering substantial mass-loss before exploding as a core-collapse
supernova (e.g., \citealt{dessart11}).  However, these values are
comparable to what was found in recent three-dimensional simulations
of neutrino-driven core collapse supernovae of He cores
\citep{muller19} -- suggesting that the progenitor of Kes 75 was
originally in a binary system which transferred most its mass to a
companion before exploding, indicative of a high initial mass.

The evolutionary model prefers a low temperature for the IC photon
field of $T = 32\,$K. This is in agreement with recent findings of
dust emitting at a temperature of $T = 33 \pm 5\,$K (for silicate
grains) by \cite{tem19}, which they conclude is most likely dust
formed by the supernova and being shock heated by the PWN.

Furthermore, since the initial spin period $P_0$ of a pulsar can be 
calculated using (e.g., \citealt{pacini73, gaensler06, slane17}):
\begin{eqnarray}
\label{eqn:period}
P_0 & = & P\left(1+\frac{t_{\rm age}}{\tau_{\rm sd}} \right)^{-\frac{1}{p-1}},
\end{eqnarray}
the similarity between the inferred spin-down timescale $\tau_{\rm
  sd}$ and $t_{\rm age}$ from this modeling suggests that $P_0 \approx
0.618 P \approx 200~{\rm ms}$ (e.g., \citealt{got00, livingstone11}).
This is considerably longer than $P_0 \approx 2~{\rm ms}$ needed to
explain is strong, spin-down inferred dipole surface magnetic field
strength $B_{\rm ns} \approx 5\times10^{13}~{\rm G}$ if it results
from an $\alpha-\Omega$ dynamo in the proto-neutron star (e.g.,
\citealt{tho93}) -- as often assumed for this and similar neutron
stars (e.g., \citealt{granot17}).  However, the high mass for the
progenitor inferred above is consistent with the notion that such
stars produce strongly magnetized neutron stars (e.g.,
\citealt{gaensler05}).

Lastly, the properties of the pulsar wind are somewhat atypical for
this class of sources.  The magnetization $\eta_{\rm B} \sim 0.07$
inferred from this modeling is $\gtrsim2\times$ higher than that
inferred from previous analyses of this system ($\eta_{\rm B} \sim
0.005$; \citealt{bucciantini11}, $\eta_{\rm B} \sim 0.008 - 0.03$;
\citealt{torres14}), though this could be the result of a limited
exploration of parameter space and reproducing a different set of
observed properties than previous work.  Of particular interest is the
value of $E_{\rm max}$.  Current theories suggest that $E_{\rm
  max,\Phi}\approx e \Phi$ (e.g., \citealt{bucciantini11}), where $e$
is the charge of the electron and $\Phi$ is the voltage near the
pulsar's polar cap (e.g., \citealt{goldreich69, bucciantini11,
  slane17}):

\begin{equation}
\begin{split}
\Phi = \sqrt{\frac{\dot{E}_{\rm psr}}{c}}& \approx  1.66\times10^{13}~{\rm statvolt} \times \frac{\rm 299.79~Volt}{\rm 1~statvolt}\\
                                         & \approx  4.99\times10^{15}~{\rm V},\\
\end{split}
\end{equation}
\medskip

\noindent or $E_{\rm max,\Phi} \approx 5~{\rm PeV}$.  While this value
isn't particularly well-constrained by our modeling, mainly due to the
lack of information concerning the MeV emission from this PWN, our
results suggest $E_{\rm max}$ agrees with \cite{bucciantini11}.

In summary, the presented X-ray observations offer a new window to
study a unique pulsar-SNR system that represents a transitional object
between the rotation-powered pulsars and the magnetars.  Continued
observations during its outburst and quiescent phases will help to
address questions related to its origin and what distinguishes \psr\
from the typical pulsar in two classes of neutron stars.

\acknowledgements 
This research was funded by the National Aeronautics and Space Administration (NASA)  \nustar\ grant 80NSSC17K0636. SSH acknowledges support from the Natural Sciences and
Engineering Research Council of Canada (NSERC) and the Canadian Space
Agency. The contributions of JDG and SMS was supported by NASA grant NNX17AL74G issued through the NNH16ZDA001N Astrophysics Data Analysis Program (ADAP).
The NuSTAR mission is a project led by the California
Institute of Technology, managed by the Jet Propulsion Laboratory, and
funded by NASA. This
research made use of the NuSTAR Data Analysis Software (NuSTARDAS)
jointly developed by the ASI Science Data Center (ASDC, Italy) and the
California Institute of Technology (USA). This research also made use
of data and software provided by the High Energy Astrophysics Science
Archive Research Center (HEASARC), which is a service of the
Astrophysics Science Division at NASA/GSFC and the High Energy
Astrophysics Division of the Smithonian Astrophysical Observatory. We
also acknowledge use of the NASA Astrophysics Data Service (ADS).

\end{document}